\documentclass[letterpaper]{mc2023}
\usepackage{tabls}
\usepackage{cites}
\usepackage{epsf}
\usepackage{appendix}
\usepackage{ragged2e}
\usepackage[top=1in, bottom=1in, left=1in, right=1in]{geometry}
\usepackage{enumitem}
\setlist[itemize]{leftmargin=*}
\usepackage{caption}
\usepackage{amsmath,amssymb,epsfig,color}
\usepackage{caption}
\usepackage{subcaption}
\usepackage{array,graphicx,multirow}

\title{Global Sensitivity Analysis in Monte Carlo Radiation Transport}
\author{%
  \textbf{K.B.~Clements$^{1,2}$, G.~Geraci$^2$, A.J.~Olson$^2$, and T.S.~Palmer$^1$}\vspace{3pt} \\
  $^1$Oregon State University  \\
  $^2$Sandia National Laboratories  \\ 
  \url{clemekay@oregonstate.edu}, \url{ggeraci@sandia.gov}, \url{aolson@sandia.gov}, \\ \url{todd.palmer@oregonstate.edu}}

\newcommand{\authorHead}{K.B.~Clements, G.~Geraci and A.J.~Olson}
\newcommand{\shortTitle}{Global Sensitivity Analysis for MC radiation transport}

\newcommand{\Var}[1]{\mathbb{V}ar\left[#1\right]}
\newcommand{\EE}[1]{\mathbb{E}\left[#1\right]}

\newcommand{\Neta}{{N_\eta}}
\newcommand{\Nxi}{{N_\xi}}

\newcommand{\EExi}[1]{\mathbb{E}_\xi\left[#1\right]}
\newcommand{\EEeta}[1]{\mathbb{E}_\eta\left[#1\right]}
\newcommand{\Vxi}[1]{\mathbb{V}ar_\xi\left[#1\right]}
\newcommand{\Veta}[1]{\mathbb{V}ar_\eta\left[#1\right]}

\newcommand{\Qpoll}{\tilde{Q}_\Neta}
\newcommand{\SigSqeta}{\sigma^2_\eta}
\newcommand{\SigSqetahat}{\hat{\sigma}^2_\eta}

\newcommand{\defin}{\triangleq}

\newcommand{\xii}{\xi_i}
\newcommand{\xini}{\xi_{\sim i}}
\newcommand{\V}[1]{\mathbb{V}ar_{#1}}
\newcommand{\E}[1]{\mathbb{E}_{#1}}
\newcommand{\Q}{Q\left( \xi \right)}

\newcommand{\eg}{\textit{e.g.}}
\newcommand{\ie}{\textit{i.e.}}

\begin{document}
\maketitle
\pagestyle{fancy} \cfoot{\thepage}
\fancypagestyle{firstpage}{\fancyhead[C]{\footnotesize{\emph{M\&C 2023 - The International Conference on Mathematics and Computational Methods Applied\\to Nuclear Science and Engineering $\cdot$ Niagara Falls, Ontario, Canada $\cdot$ August 13 -- 17, 2023}}}
\cfoot{}}
\thispagestyle{firstpage}
\fancyhead[CE]{{\scriptsize \authorHead}}
\fancyhead[CO]{{\scriptsize \shortTitle}}
\justify 
\parskip 6pt plus 1 pt minus 1 pt

\begin{abstract}
We consider \emph{Global Sensitivity Analysis} (GSA) for Monte Carlo (MC) radiation transport (RT) applications. GSA is usually combined with Uncertainty Quantification (UQ), where the latter (among other goals) quantifies
the variability of a model output in the presence of uncertain inputs and the former attributes this variability to the inputs. 
The additional noise inherent to MC RT solvers due to the finite number of particle histories presents an additional challenge to GSA and UQ, which are well-established for deterministic solvers. 
In this contribution, we apply variance deconvolution to Saltelli's method to address MC RT solver noise without having to over-resolve the MC RT simulation.
\end{abstract}
\vspace{6pt}

\section{INTRODUCTION} 
\label{sec:intro}
Global sensitivity analysis (GSA) aims to apportion, or divide and allocate, variability in model output to different sources of uncertainty in model input~\cite{Saltellibook}. GSA is useful to understand the relative importance of each of a model's uncertain inputs, and their interactions with one another, to the behavior of model output. It is typically paired with uncertainty quantification (UQ), which deals with characterizing and propagating uncertainty sources through computational models. For an exhaustive introduction to GSA in the scientific computing context, see Saltelli's book~\cite{Saltellibook}. 
This work focuses on sampling-based GSA applied to stochastic solvers, specifically Monte Carlo radiation transport (MC RT) solvers. Typically, UQ and GSA assume that the computational model itself is deterministic, \ie, that given identical inputs, the model will produce identical outputs. From this assumption, it follows that any output variability characterized by UQ or apportioned by GSA is a result of some uncertain input to the solver, not variability inherent to the solver itself. Despite the abundant literature produced on GSA over the last few decades, there is a gap in the quantification and control of the intrinsic randomness introduced by non-deterministic solvers. 

Stochastic solvers are widely used and important for many applications depending on the information needed by the user, the problem space, and the complexity of the modeled system. Unlike deterministic methods, which require phase-space discretization and approximate solutions to continuous equations, MC RT methods are event based and can faithfully model complex physics. This makes MC RT methods well-suited to solve, for example, complicated three-dimensional, time-dependent problems~\cite{mcnp}. However, results from MC RT solvers are always approximate, constrained by the finite number of particle histories that can be used in a simulation. While it is certainly possible to apply UQ and GSA to stochastic solvers, this invalidates the assumption that output uncertainty can be analyzed solely in the context of input uncertainty. Statistical analysis can be considered ``polluted'' by the variability introduced by the solver itself. It is possible to rigorously show~\cite{ClementsCSRI2021,ClementsANS2022} that the stochastic solver increases the observed model output variance, possibly causing an analyst to over-estimate the model's response to an uncertain input. 
A brute-force method to address this complication and ``de-pollute'' statistics of interest is to over-resolve the stochastic solver, \eg, increase the number of particles in the simulation, until the solver variance is rendered negligible compared to the effects of the uncertain inputs~\cite{MCMC-paper}. Resolving stochastic models to this extent is already computationally expensive and folding that into the UQ and GSA workflow, which requires repeated evaluation of numerical codes, increases the computational expense to the point of intractability. Our goal is to gain an understanding of how GSA can be performed in the context of MC RT solvers by explicitly accounting for the stochastic variability they introduce. 



This work builds on a recently-derived \emph{variance deconvolution} approach~\cite{ClementsCSRI2021,ClementsANS2022}. We have introduced variance deconvolution to quantify the variance contribution from a stochastic solver and effectively remove it from the total polluted variance, accurately estimating the desired variance induced by uncertain input parameters (referred to from here as \emph{parametric variance}). This is far more cost effective than the brute-force approach, and uses an unbiased estimator for the variance introduced by the solver and for the parametric variance. We apply this variance deconvolution UQ workflow here to MC RT problems, but as we will show, the method is not specific to radiation transport and is widely applicable to stochastic solvers. Also in recent work, we integrated variance deconvolution in sampling-based GSA for stochastic media~\cite{OlsonANSWinter} and surrogate~\cite{GeraciMC2021,GeraciMC2023} approaches. 
Detailed derivation and analysis for UQ with variance deconvolution is available in~\cite{ClementsANS2022}, and is summarized below.
In this paper, we apply variance deconvolution to a general GSA case and compare its performance to the straightforward application of Saltelli's method, without any particular correction for the solver's noise.

\section{Global Sensitivity Analysis: Background Theory} 
\label{sec:theory}
We consider a generic QoI $Q=\Q$, which expresses a mapping from the vector of $d$ uncertain input parameters $\xi \in \Xi \subset \mathbb{R}^d$, with joint probability density function (PDF) $p(\xi)$, to scalar $Q$. In standard UQ, we are concerned with estimating statistics for $Q$ with respect to the input parameters, \textit{e.g.} moments like the mean and variance:
\begin{equation}
\label{eq:UQmoments}
 \EExi{Q} \defin \int_\Xi Q(\xi) p(\xi) d\xi \quad \mathrm{and} \quad 
 \Vxi{Q} \defin \int_\Xi \left( Q(\xi) - \EExi{Q} \right)^2 p(\xi) d\xi.
\end{equation}
In this work, we consider variance-based GSA\footnote{We limit ourselves here to variance-based strategies, although other approaches are also possible~\cite{Owen2014,Geraci2016}.}, quantifying the uncertainty of model output by studying how each parameter (or group of parameters) $\xii$ affects the output's variance. We start by considering that $\xii$ is fixed to some value in its PDF $\xii^*$. To compute the mean of $Q$ conditional on $\xii=\xii^*$, we take the expected value of $Q$ over all parameters \textit{except} $\xii$, denoted $\xini$. 
The conditional variance 
over all possible values of $\xii^*$, $
    \V{\xii} \Bigl[ \E{\xini}\bigl[ Q \mid \xii \bigr] \Bigr] \defin \mathbb{V}_i $,
is known as the first-order effect of $\xii$ on $Q$, a measure of the variance introduced by parameter $\xii$. To simplify notation, we write this as $\Var{\EE{Q \mid \xii}}$, where the parameters of integration can be assumed from the fixed parameter. We can also consider higher order effects, known as interaction effects, which captures that $Q$'s response to a set of parameters cannot be fully described by the sum of their individual first-order effects. For example, the second-order effect of the pair $\left(\xii,\xi_j\right)$ can be written using Sobol's decomposition~\cite{Saltelli} by removing their individual first-order effects from their joint effect, 
$ \mathbb{V}_{ij} = \Var{\EE{Q \mid \xi_i, \xi_j}} - \mathbb{V}_i - \mathbb{V}_j$. 
While $\mathbb{V}_i$ is a measure of the effect of $\xii$ on $Q$, the second-order effect $\mathbb{V}_{ij}$ is a measure of the effect of the interaction between $\xii$ and $\xi_j$ on $Q$.

Sensitivity indices, sometimes referred to as Sobol' indices (SI), provide a measure of how important a parameter (or set of parameters) is in contributing to the overall variance. This sensitivity can range between 0 and 1, where importance increases as a SI approaches 1.
A SI can be computed for any of a parameter's arbitrary-order effects. We are typically interested in computing the first-order SI $S_i$ and total SI $S_{Ti}$,
\begin{equation}\label{eq:si}
    S_i = \frac{\V{\xii} \Bigl[ \E{\xini}\bigl[ Q \mid \xii \bigr] \Bigr]}{\Vxi{Q}}
    , \quad \quad
    S_{Ti} = 1 - \frac{\V{\xini} \Bigl[ \E{\xii}\bigl[ Q \mid \xini \bigr] \Bigr]}{\Vxi{Q}} ,
\end{equation} 
where total SI accounts for the individual effect of $\xii$ and all of its interaction effects. For a model with three uncertain input factors, the total effect SI of $\xi_1$ is the sum of its first-order, second-order, and third-order SIs, $S_{T1} = S_1 + S_{12} + S_{13} + S_{123}$.

\subsection{Saltelli's method}
Saltelli introduced a widely used sampling method~\cite{Saltellibook} that provides the benchmark for any subsequent GSA development, which we briefly summarize here. Assuming $d$ random inputs and $N$ sampling realizations, Saltelli's algorithm reads as follows:
\begin{enumerate}
    \item Define two $(N,d)$ matrices, $A$ and $B$, which contain independent input samples. 
    \begin{equation}\label{eq:ABmatrices}
        A = \begin{bmatrix}
        \xi_1^{(1)} & \cdots & \xi_i^{(1)}   & \cdots & \xi_d^{(1)}  \\
        \vdots      &        & \ddots        &        & \vdots       \\
        \xi_1^{(N)} & \cdots & \xi_i^{(N)}   & \cdots & \xi_d^{(N)}  \\
        \end{bmatrix} 
        , \quad \quad
        B = \begin{bmatrix}
        \xi_{d+1}^{(1)} & \cdots & \xi_{d+i}^{(1)}   & \cdots & \xi_{2d}^{(1)}   \\
        \vdots          &        & \ddots            &        & \vdots           \\
        \xi_{d+1}^{(N)} & \cdots & \xi_{d+i}^{(N)}   & \cdots & \xi_{2d}^{(N)}   \\
        \end{bmatrix} .
    \end{equation}
    \item For each $i$th random input, define a matrix $C_i$ using all columns of $B$ except for the $i$th column, which comes from $A$. 
    \begin{equation}\label{eq:Cmatrix}
        C_i = \begin{bmatrix}
        \xi_{d+1}^{(1)} & \cdots & \xi_{i}^{(1)} & \cdots & \xi_{2d}^{(1)}    \\
        \vdots          &        & \ddots        &        & \vdots            \\
        \xi_{d+1}^{(N)} & \cdots & \xi_{i}^{(N)} & \cdots & \xi_{2d}^{(N)}    \\
        \end{bmatrix} .
    \end{equation}
    \item Compute model output for $A$, $B$, and all $C_i$ to obtain vectors of model output $y$ of dimension $(N,1)$. 
    \item Estimate the first-order and total sensitivity indices via sampling:
    \begin{align}\label{eq:sampling-si}
        S_i = \frac{ \Var{\EE{Q \mid \xii}} }{\Vxi{Q}} 
        &\approx \frac{\frac{1}{N}\sum_{j=1}^N y_A^{(j)}y_{C_i}^{(j)} - \left( \frac{1}{N}\sum_{j=1}^N y_A^{(j)}\right)^2}
               {\frac{1}{N}\sum_{j=1}^N \left(y_A^{(j)}\right)^2 - \left( \frac{1}{N}\sum_{j=1}^N y_A^{(j)}\right)^2} , \\
    \label{eq:sampling-sti}
        S_{T_i} = 1 - \frac{ \Var{\EE{Q \mid \xini}} }{\Vxi{Q}}
        &\approx 1 - \frac{ \frac{1}{N}\sum_{j=1}^N y_B^{(j)}y_{C_i}^{(j)} - \left( \frac{1}{N}\sum_{j=1}^N y_A^{(j)}\right)^2 }
               {\frac{1}{N}\sum_{j=1}^N \left(y_A^{(j)}\right)^2 - \left( \frac{1}{N}\sum_{j=1}^N y_A^{(j)}\right)^2} .
    \end{align}
\end{enumerate}

While in this paper we only consider the baseline Saltelli method as in~\cite{Saltellibook}, a number of modifications and extensions have been made to the method, for example as discussed in~\cite{Saltelli}.

\section{Computing Sobol' indices with stochastic solvers}
\label{sec:sampling-gsa}
%
The Saltelli method is well-established when the computational model is deterministic. We propose a modified application of the Saltelli method for use with stochastic computational models by applying variance deconvolution. 

When the computational model is a stochastic solver, the QoI $Q$ can only be approximated by averaging a finite number of elementary event realizations $f$ (see~\cite{ClementsANS2022} for details). For instance, in MC RT applications, we indicate with $f$ an event resulting from a single particle history, and approximate $Q$ using $\Neta$ particle histories:
\begin{equation}\label{eq:qpoll}
 Q(\xi) \defin \EEeta{ f(\xi,\eta) } \approx \frac{1}{\Neta} \sum_{j=1}^{\Neta} f( \xi, \eta^{(j)} ) \defin \Qpoll(\xi).
\end{equation}
The additional variable $\eta$ is introduced only to notionally represent the randomness in a MC RT solver. In practice $\eta$, unlike $\xi$, is neither controlled nor assumed to be known, and merely reflects that even for identical systems defined by the same $\xi$, individual particle histories will follow different trajectories. 
Because $Q$ is approximated by $\Qpoll$, parametric variance is not directly accessible. With the variance deconvolution approach~\cite{ClementsANS2022}, we approximate the parametric variance $\Vxi{Q}$ from observable quantities via Eq.~\eqref{eq:var-deconv}, where $\Var{\Qpoll}$ represents the total variance (polluted by the MC RT noise) and $\EExi{ \SigSqeta }$ represents the average contribution from the solver's stochasticity $
\SigSqeta\defin \Veta{ f }$ 
\begin{equation}\label{eq:var-deconv}
 \Vxi{Q} = \Var{\Qpoll} - \frac{\EExi{ \SigSqeta }}{\Neta}.
\end{equation}

In~\cite{OlsonANSWinter}, we applied this variance deconvolution strategy to GSA in the case where the QoI was the conditional expectation of $Q$ over stochastic media realizations. Here, we focus on the general case, wherein we desire to compute first-order and total SIs (Eq.~\eqref{eq:si}) for the QoI $Q$ but can only access 
$\Qpoll$. 
We develop an expression for the first-order effect $\Var{\EE{Q \mid \xii}}$ by first applying the law of total variance to the polluted total variance, 
\begin{equation}\label{eq:one}
    \Var{\Qpoll} = \V{\xii}\biggl[ \E{\xini,\eta} \left[ \Qpoll \right] \biggr] + \E{\xii}\biggl[ \V{\xini,\eta} \left[ \Qpoll \right] \biggr] .
\end{equation}
We apply variance deconvolution and the law of total variance as needed and, after a few manipulations, arrive at an expression for the first-order effect of $\xii$,
%
\begin{equation}\label{eq:first-order-deconv}
    \V{\xii}\biggl[ \E{\xini} \bigl[ Q \bigr] \biggr] = \Var{\Qpoll} - \E{\xii}\biggl[\V{\xini} \bigl[ \Qpoll \bigr]\biggr] ,
\end{equation}
where we have refrained from including the full derivation details in the interest of space. It follows that the first-order effect of $\xini$, needed to compute the total SI, can be written
\begin{equation}\label{eq:total-deconv}
    \V{\xini}\biggl[ \E{\xii} \bigl[ Q \bigr] \biggr] = \Var{\Qpoll} - \E{\xini}\biggl[\V{\xii} \bigl[ \Qpoll \bigr]\biggr] .
\end{equation}

When applying variance deconvolution to UQ~\cite{ClementsANS2022}, we computed the parametric variance by removing the average solver variance from the total polluted variance. Applied here for GSA, however, we do not need to explicitly compute the average solver variance to compute the parametric conditional variances. Instead, we do so by removing the polluted conditional means, \eg, $\EE{\Var{\Qpoll | \xii}}$ from the total polluted variance. 

\subsection{Modifying Saltelli's method for stochastic solvers}
We use the sampling scheme from Saltelli's method to compute sample estimates
of Eqs.~\eqref{eq:first-order-deconv} and~\eqref{eq:total-deconv}, analogous to the sampling estimators in Eqs.~\eqref{eq:sampling-si} and~\eqref{eq:sampling-sti} for the deterministic-solver case. We consider matrices of random numbers $A$, $B$, and $C_i$ as defined in Eqs.~\eqref{eq:ABmatrices} and~\eqref{eq:Cmatrix}, and output vectors $\tilde{y}_A$, $\tilde{y}_B$, and $\tilde{y}_{C_i}$ that contain outputs of an MC RT solver $\Qpoll \left( \xi_1, \ldots, \xi_d\right)$. The only new information to collect for our modification is a vector of solver variances 
\begin{equation}
    \tilde{y}_{\SigSqeta,A} \defin \begin{bmatrix}
            \SigSqetahat\left( \xi_1^{(1)}, \cdots, \xi_i^{(1)}, \cdots, \xi_d^{(1)}\right) \\
            \SigSqetahat\left( \xi_1^{(2)}, \cdots, \xi_i^{(2)}, \cdots, \xi_d^{(2)}\right) \\
            \vdots \\
            \SigSqetahat\left( \xi_1^{(N)}, \cdots, \xi_i^{(N)}, \cdots, \xi_d^{(N)}\right) \\
            \end{bmatrix}, \;
\end{equation}
where $\SigSqetahat$ is the sampling estimator for $\SigSqeta$, 
\begin{equation}
    \SigSqetahat\left(\xi^{(k)} \right) \defin \frac{1}{\Neta}\sum_{j=1}^\Neta \left( f\left( \xi^{(k)},\eta^{(j)}\right) - \Qpoll\left(\xi^{(k)}\right) \right)^2 .
\end{equation}

We define sampling estimator counterparts for the terms in Eq.~\eqref{eq:first-order-deconv} and~\eqref{eq:total-deconv}. The total polluted variance,
\begin{equation}
    \Var{\Qpoll} \approx \tilde{S}^2 \defin \frac{1}{N-1} \sum_{j=1}^N \left( \tilde{y}_A^{(j)} - \sum_{k=1}^N \tilde{y}_A^{(k)}  \right)^2 ,
\end{equation}
is used to estimate the parametric variance,
\begin{equation}
    \Vxi{Q} \approx S^2 \defin \tilde{S}^2 - \frac{1}{\Neta \Nxi} \sum_{j=1}^N \tilde{y}_{\SigSqeta,A}^{(j)} .
\end{equation}

We estimate the conditional variances as
\begin{equation}
    \V{\xini} \bigl[ \Qpoll^{(j)} \bigr] \approx \left(\tilde{S}^{2}_{\xini} \right) ^{(j)} \defin \left(\tilde{y}_A^{(j)}\right)^2 + \left(\tilde{y}_{C_i}^{(j)}\right)^2 - \frac{\left( \tilde{y}_A^{(j)} + \tilde{y}_{C_i}^{(j)} \right)^2}{2},
\end{equation}
\begin{equation}
    \V{\xii} \bigl[ \Qpoll^{(j)} \bigr] \approx \left(\tilde{S}^{2}_{\xii} \right) ^{(j)} \defin \left(\tilde{y}_B^{(j)}\right)^2 + \left(\tilde{y}_{C_i}^{(j)}\right)^2 - \frac{\left( \tilde{y}_B^{(j)} + \tilde{y}_{C_i}^{(j)} \right)^2}{2}
\end{equation}
such that the sample estimators for first-order and total SI using stochastic solvers are:
\begin{equation}
    S_i = \frac{ \Var{\EE{Q \mid \xii}} }{\Vxi{Q}} \approx 
    \frac{ \tilde{S}^2 - \frac{1}{N} \sum_{j=1}^N  \left(\tilde{S}^{2}_{\xini} \right) ^{(j)}}{S^2}
\end{equation}
\begin{equation}
    S_{T_i} = 1 - \frac{ \Var{\EE{Q \mid \xini}} }{\Vxi{Q}} \approx 
    1 - \frac{\tilde{S}^2 - \frac{1}{N} \sum_{j=1}^N  \left(\tilde{S}^{2}_{\xii} \right) ^{(j)}}{S^2} .
\end{equation}


\section{Results} 
\label{sec:test_case}
As a test radiation transport problem, we consider a neutral-particle, attenuation-only, mono-energetic steady-state radiation transport problem. A beam source of magnitude one is incident on a 1D slab of length 3 that is separated into three material regions. The problem QoI $\Q$ is transmittance through the slab. We introduce six uniformly distributed uncertain parameters, grouped into three uncertainty sources of interest:
$1)$ Cosine of beam-source incidence angle $\mu \sim \mathcal{U}\left[0.6,1.0\right]$;
$2)$ Boundary locations between material regions $x_1 \sim \mathcal{U}\left[0.3,1.7\right]$ and $x_2 \sim \mathcal{U}\left[1.7,2.3\right]$;
and $3)$ Total cross sections of the slab materials $\Sigma_{t,1} \sim \mathcal{U}\left[0.1,0.9\right], \Sigma_{t,2} \sim \mathcal{U}\left[0.2,0.4\right], \Sigma_{t,3} \sim \mathcal{U}\left[0.07,1.03\right]$.

To investigate our variance deconvolution modification of Saltelli's method (denoted Saltelli-VarD for brevity), we compute first-order and total SIs for the three groups of parametric uncertainty with both Saltelli-VarD and the straightforward Saltelli method. For benchmark reference solutions, we solve for SIs using Saltelli's method with $N=10^8$ sample realizations, computing transmittance analytically with optical thickness.

\begin{figure}
     \centering
     \begin{subfigure}[b]{0.49\textwidth}
         \centering
         \includegraphics[width=\textwidth]{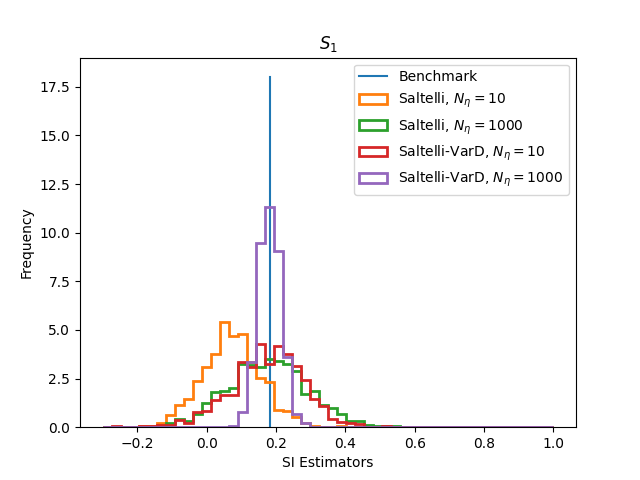}
         \caption{First-order SI for cosine of beam incident angle}
         \label{fig:S1}
     \end{subfigure}
     \hfill
     \begin{subfigure}[b]{0.49\textwidth}
         \centering
         \includegraphics[width=\textwidth]{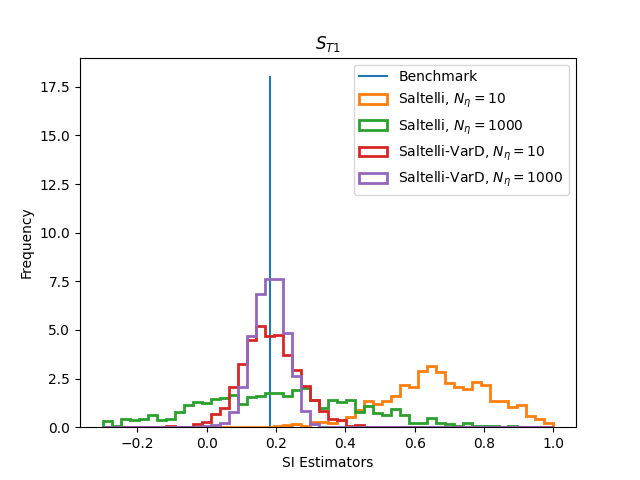}
         \caption{Total SI for cosine of beam incident angle}
         \label{fig:ST1}
     \end{subfigure}
     \hfill
     \begin{subfigure}[b]{0.49\textwidth}
         \centering
         \includegraphics[width=\textwidth]{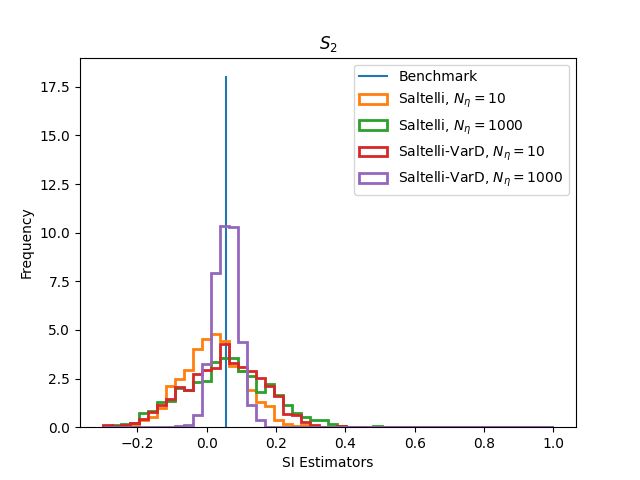}
         \caption{First-order SI for material boundary location}
         \label{fig:S2}
     \end{subfigure}
     \hfill
     \begin{subfigure}[b]{0.49\textwidth}
         \centering
         \includegraphics[width=\textwidth]{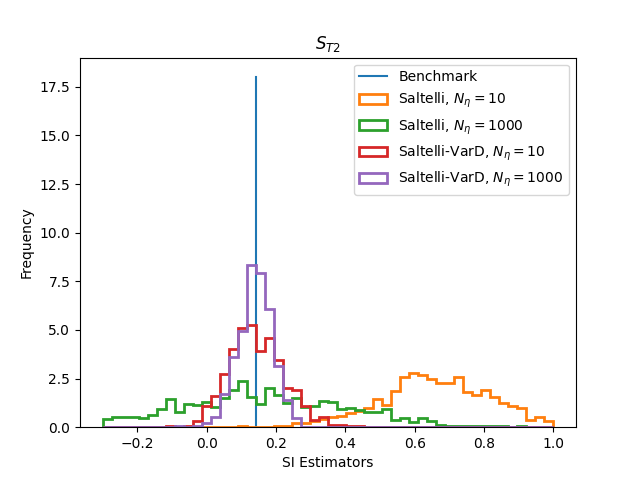}
         \caption{Total SI for material boundary location}
         \label{fig:ST2}
     \end{subfigure}
     \hfill
     \begin{subfigure}[b]{0.49\textwidth}
         \centering
         \includegraphics[width=\textwidth]{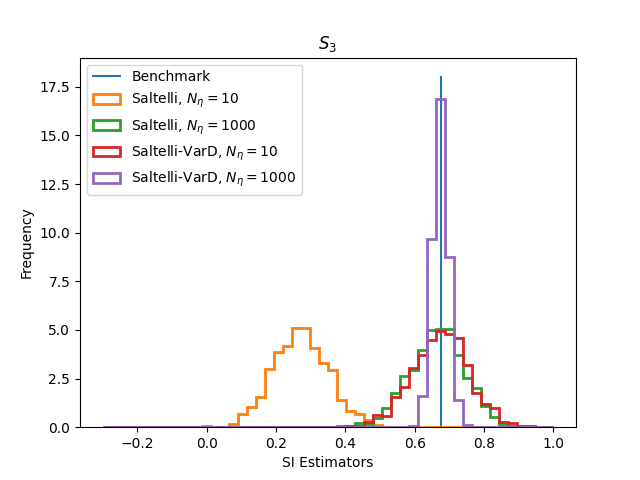}
         \caption{First-order SI for total cross section}
         \label{fig:S3}
     \end{subfigure}
     \hfill
     \begin{subfigure}[b]{0.49\textwidth}
         \centering
         \includegraphics[width=\textwidth]{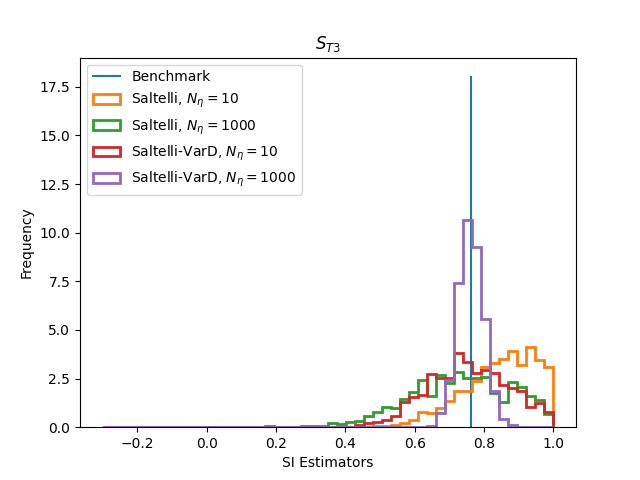}
         \caption{Total SI for total cross section}
         \label{fig:ST3}
     \end{subfigure}
        \caption{First-order and total sensitivity indices for the three groups of parametric uncertainty. Comparing using the straightforward Saltelli approach and Saltelli with variance deconvolution (Saltelli-VarD) over 1000 repetitions. MC RT simulations performed with Sandia National Laboratories research code PlaybookMC. }
        \label{fig:results}
\end{figure}

GSA is performed using $N=\Nxi$ sample realizations, where each model realization is a MC RT simulation using $\Neta$ particle histories. We perform two GSA tests, one using $(\Nxi=1000, \Neta=10)$, and another using $(\Nxi=1000, \Neta=1000)$. We repeat this numerical experiment 1000 times to construct histograms of estimator output, shown in Figure~\ref{fig:results}. 
Across all of the histograms in Figure~\ref{fig:results}, we can see that applying Saltelli's method, developed for deterministic solvers, to a stochastic solver does indeed produce biased results for first-order and total SI compared to the benchmark result. This is most visible in $S_3$ and $S_{T1}$, Figures~\ref{fig:S3} and~\ref{fig:ST1}. 
For every SI, we can see that the Saltelli-VarD result is unbiased compared to the benchmark solution, corroborating our theoretical finding that the conditional variance is accessible using variance deconvolution. In using Saltelli's method with a stochastic solver, one could increase the number of particle histories per simulation to drive down the MC RT variance. We can see the effect of this approach looking at the $\Neta=1000$ case, where we see that all of the results are converging to the benchmark mean, confirming that over-resolving the MC RT simulation will eventually drive down the solver variance and cause the bias term to approach 0.
However, in every case, the Saltelli-VarD results with $\Neta=10$ are distributed similarly to Saltelli's method with $\Neta=1000$.
For this particular example problem, we observe that Saltelli-VarD produces results comparable to Saltelli's method with $100 \times$ fewer particles per sample.
Comparing Saltelli-VarD's $\Neta=1000$ results to Saltelli's $\Neta=1000$ results, we see that Saltelli-VarD achieves a much tighter distribution around the benchmark solution for all first-order and total SIs.

To quantify these effects, we compute the mean-squared error (MSE) for both methods compared to the benchmark result.
Eq.~\eqref{eq:mse} shows MSE of estimator value $\hat{X}$ with respect to a known result $X$, from which we can see that MSE captures both the variance and the bias of the estimator,
\begin{equation}\label{eq:mse}
        MSE \left[ \hat{X} \right] = \EE{\left( \hat{X} - X \right)^2} = \Var{\hat{X}} + Bias^2 \left[ \hat{X},X \right] .
\end{equation}

In Table~\ref{tab:results-mean} we compare the SIs computed with Saltelli and Saltelli-VarD (averaged over 1000 repetitions) and the benchmark SI values. Using $\Neta=10$, Saltelli-VarD well-approximates the benchmark result, while Saltelli's method has statistically significant deviation from the benchmark results.
In Table~\ref{tab:results-mse}, we report the variance, bias, and MSE of each SI from both methods. For Saltelli's method, going from $\Neta=10$ to $\Neta=1000$ does reduce the bias term, corresponding to how one might hope to over-resolve the MC RT solution by increasing the number of particle histories. This comparison between $\Neta=10$ and $\Neta=1000$ is for a simple 1D slab and attenuation-only physics, and the computational cost of MC RT resolution will only increase with problem complexity.
While we do see non-zero bias results for Saltelli-VarD, the standard deviation indicates zero-bias is within a $1\sigma$ confidence interval, corroborating our theoretical finding that the Saltelli-VarD approach provides an unbiased estimate of SIs.

\begin{table}[ht]
    \centering
    \caption{Comparing Saltelli's method to Saltelli-VarD for computing first-order and total SIs when using a MC RT solver. Saltelli and Saltelli-VarD results averaged over 1000 repetitions, using $\Nxi=1000$ for every case. Results indicate mean(std dev).}
    \begin{tabular}{| c || l || l | l || l | l |} \hline
         & \multicolumn{1}{c||}{Benchmark} & \multicolumn{2}{c||}{Saltelli} & \multicolumn{2}{c|}{Saltelli-VarD} \\
        $\Neta$ & & \multicolumn{1}{c|}{$10$} & \multicolumn{1}{c||}{$1000$} & \multicolumn{1}{c|}{$10$} & \multicolumn{1}{c|}{$1000$} \\ \hline \rule{0pt}{13pt}
        $S_1$    & 0.1824(4)  & 0.07(8) & 0.2(1)  & 0.2(1)  & 0.18(3) \\
        $S_2$    & 0.0549(4)  & 0.02(9) & 0.1(1)  & 0.0(1)  & 0.06(3) \\
        $S_3$    & 0.6752(2) & 0.27(8) & 0.66(8) & 0.67(8) & 0.67(2) \\
        $S_{T1}$ & 0.1835(8)  & 0.7(2)  & 0.2(2)  & 0.18(8) & 0.18(5) \\
        $S_{T2}$ & 0.1424(8)  & 0.7(2)  & 0.2(3)  & 0.14(8) & 0.14(5) \\
        $S_{T3}$ & 0.7623(7)  & 0.9(1)  & 0.8(2)  & 0.8(1)  & 0.76(4) \\ \hline
    \end{tabular}
    \label{tab:results-mean}
\end{table}

\begin{table}[ht]
    \centering
    \caption{MSE, variance, and bias of Saltelli and Saltelli-VarD methods. Comparing performance using $\Neta=10$ to using $\Neta=1000$ with a constant $\Nxi=1000$, over $1000$ repetitions.}
    \begin{tabular}{|c| c || r | r | r | r | r | r |} \hline
        \rule{0pt}{13pt} & & \multicolumn{2}{c|}{MSE} & \multicolumn{2}{c|}{Var} & \multicolumn{2}{c|}{Bias} \\
        & $\Neta$ & \multicolumn{1}{c|}{$10$} & \multicolumn{1}{c|}{$1000$} & \multicolumn{1}{c|}{$10$} & \multicolumn{1}{c|}{$1000$} & \multicolumn{1}{c|}{$10$} & \multicolumn{1}{c |}{$1000$} \\ \hline \rule{0pt}{13pt}
        {\multirow{6}{*}{\rotatebox[origin=c]{90}{Saltelli}}} & $S_1$    & 0.0194 & 0.0130 & 0.0068 & 0.0129 & -0.1121 & -0.0077 \\
        & $S_2$    & 0.0089 & 0.0155 & 0.0075 & 0.0155 & -0.0368 & -0.0006 \\
        & $S_3$    & 0.1693 & 0.0064 & 0.0062 & 0.0063 & -0.4038 & -0.0111 \\
        & $S_{T1}$ & 0.2626 & 0.0562 & 0.0233 & 0.0560 & 0.4892  & 0.0151  \\
        & $S_{T2}$ & 0.2914 & 0.0639 & 0.0252 & 0.0638 & 0.5160  & 0.0115  \\
        & $S_{T3}$ & 0.0396 & 0.0313 & 0.0189 & 0.0313 & 0.1437  & 0.0047  \\ \hline \hline \rule{0pt}{13pt}
        & & \multicolumn{2}{c|}{MSE} & \multicolumn{2}{c|}{Var} & \multicolumn{2}{c|}{Bias} \\
        & $\Neta$ & \multicolumn{1}{c|}{$10$} & \multicolumn{1}{c|}{$1000$} & \multicolumn{1}{c|}{$10$} & \multicolumn{1}{c|}{$1000$} & \multicolumn{1}{c|}{$10$} & \multicolumn{1}{c |}{$1000$} \\ \hline \rule{0pt}{13pt}
        {\multirow{6}{*}{\rotatebox[origin=c]{90}{Saltelli-VarD}}} & $S_1$    & 0.0109 & 0.0010 & 0.0108 & 0.0010 & -0.0099 & -0.0013 \\
        & $S_2$    & 0.0124 & 0.0012 & 0.0123 & 0.0012 & -0.0091 & 0.0008 \\
        & $S_3$    & 0.0066 & 0.0005 & 0.0066 & 0.0005 & -0.0007 & -0.0010 \\
        & $S_{T1}$ & 0.0064 & 0.0023 & 0.0064 & 0.0023 & 0.0004  & -0.0015 \\
        & $S_{T2}$ & 0.0063 & 0.0024 & 0.0063 & 0.0024 & 0.0016  & -0.0024 \\
        & $S_{T3}$ & 0.0187 & 0.0013 & 0.0186 & 0.0013 & 0.0104  & 0.0003  \\ \hline
    \end{tabular}
    \label{tab:results-mse}
\end{table}

%


%

\section{CONCLUSIONS}
The Saltelli method is a well-defined approach for global sensitivity analysis (GSA), but assumes that analysis is performed using a deterministic solver.
In this paper, we consider the effects on GSA results of using a stochastic solver, namely a Monte Carlo radiation transport solver.
We have incorporated our previously-developed variance deconvolution approach~\cite{ClementsANS2022} to the Saltelli method for GSA and compared its performance to the unmodified approach. 
Applied to a test 1D radiation transport problem with three independent sources of parametric variance, our approach accurately estimated first-order and total sensitivity indices for significantly less computational cost than the unmodified Saltelli method and, for the same computational cost, out-performed the unmodified Saltelli method in terms of accuracy.

%

\section*{ACKNOWLEDGMENTS}
This article has been authored by an employee of National Technology \& Engineering Solutions of Sandia, LLC under Contract No. DE-NA0003525 with the U.S. Department of Energy (DOE). The employee owns all right, title and interest in and to the article and is solely responsible for its contents. The United States Government retains and the publisher, by accepting the article for publication, acknowledges that the United States Government retains a non-exclusive, paid-up, irrevocable, world-wide license to publish or reproduce the published form of this article or allow others to do so, for United States Government purposes. The DOE will provide public access to these results of federally sponsored research in accordance with the DOE Public Access Plan \texttt{https://www.energy.gov/downloads/doe-public-access-plan}.

This work was supported by the Center for Exascale Monte-Carlo Neutron Transport (CEMeNT), a PSAAP-III project funded by the Department of Energy, grant number DE-NA003967.

\newif\ifusebibtex
\usebibtextrue

\ifusebibtex
\setlength{\baselineskip}{12pt}
\bibliographystyle{mc2023}
\bibliography{mc2023}

\begin{thebibliography}{10}
\newcommand{\enquote}[1]{``#1''}
\providecommand{\url}[1]{\texttt{#1}}
\providecommand{\urlprefix}{URL }

\bibitem{Saltellibook}
A.~Saltelli et~al.
\newblock \emph{Global Sensitivity Analysis: The primer}.
\newblock John Wiley \& Sons (2008).

\bibitem{mcnp}
Los Alamos National Laboratory.
\newblock \emph{MCNP - A General Monte Carlo N-Particle Transport Code, Version
  5} (2008).

\bibitem{ClementsCSRI2021}
K.~Clements, G.~Geraci, and A.~J. Olson.
\newblock \enquote{A Variance Deconvolution Approach to Sampling Uncertainty
  Quantification for Monte Carlo Radiation Transport Solvers.}
\newblock In \emph{Computer Science Research Institute Summer Proceedings
  2021}, Technical Report SAND2022-0653R, pp. 293--307 (2021).
\newblock \url{https://cs.sandia.gov/summerproceedings/CCR2021.html}.

\bibitem{ClementsANS2022}
K.~Clements, G.~Geraci, and A.~J. Olson.
\newblock \enquote{Numerical investigation on the performance of a variance
  deconvolution estimator.}
\newblock \emph{Trans Am Nucl Soc}, \textbf{volume 126}, pp. 344--347 (2022).

\bibitem{MCMC-paper}
E.~Cho and M.~J. Cho.
\newblock \enquote{Variance of Sample Variance.}
\newblock \emph{Proceedings of the Survey Research Methods Section}, pp.
  1291--1293 (2008).

\bibitem{OlsonANSWinter}
A.~J. Olson, K.~Clements, and J.~Petticrew.
\newblock \enquote{A sampling-based approach to solve Sobol’ Indices using
  variance deconvolution for arbitrary uncertainty distributions.}
\newblock \emph{Trans Am Nucl Soc}, \textbf{volume 127}, pp. 450--453 (2022).

\bibitem{GeraciMC2021}
G.~Geraci and A.~J. Olson.
\newblock \enquote{Impact of sampling strategies in the polynomial chaos
  surrogate construction for Monte Carlo transport applications.}
\newblock In \emph{Proceedings of ANS M\&C}, pp. 76--86 (2021).

\bibitem{GeraciMC2023}
G.~Geraci, K.~Clements, and A.~J. Olson.
\newblock \enquote{A Polynomial Chaos Approach for Uncertainty Quantification
  of Monte Carlo Transport Codes.}
\newblock In \emph{Proceedings of ANS M\&C} (2023).

\bibitem{Owen2014}
A.~B. Owen, J.~Dick, and S.~Chen.
\newblock \enquote{{Higher order Sobol' indices}.}
\newblock \emph{Information and Inference: A Journal of the IMA},
  \textbf{volume~3}(1), pp. 59--81 (2014).
\newblock \urlprefix\url{https://doi.org/10.1093/imaiai/iau001}.

\bibitem{Geraci2016}
G.~Geraci, P.~Congedo, R.~Abgrall, and G.~Iaccarino.
\newblock \enquote{High-order statistics in global sensitivity analysis:
  Decomposition and model reduction.}
\newblock \emph{Computer Methods in Applied Mechanics and Engineering},
  \textbf{volume 301}, pp. 80--115 (2016).
\newblock
  \urlprefix\url{https://www.sciencedirect.com/science/article/pii/S0045782515004284}.

\bibitem{Saltelli}
A.~Saltelli et~al.
\newblock \enquote{Variance based sensitivity analysis of model output. Design
  and estimator for the total sensitivity index.}
\newblock \emph{Computer Physics Communications}, \textbf{volume 181}, pp.
  259--270 (2010).

\end{thebibliography}


\begin{thebibliography}{300}
\bibitem{journal} B. Author(s), ``Title, using capitalization'' \emph{Journal Name in Italic}, 
  \textbf{Volume in Bold}, pp. 34-89 (20xx).
\bibitem{proc_paper} C. D. Author(s), ``Article Title,'' \emph{Proceedings of
  Meeting in Italic}, Location, Dates of Meeting, Vol. n, pp. 134-156 
  (20xx).
\bibitem{book} E. F. Author, \emph{Book Title in Italic}, Publisher, City \&
  Country (20xx). 
\bibitem{website} ``Canadian SMR Roadmap,'' \\
  \url{https://smrroadmap.ca/wp-content/uploads/2018/12/Technology-WG.pdf} (2018).
\end{thebibliography}
\else
\setlength{\baselineskip}{12pt}

\fi


\end{document}